\documentclass[prl,twocolumn,showpacs]{revtex4}
\usepackage{amsfonts}
\usepackage{amsmath}
\usepackage{graphicx}
\usepackage{amssymb}

\def\comment#1{}

\def\mn#1{*\marginpar{*\tiny{#1}}}
\def\mn#1{}

\usepackage{color}
\usepackage{ulem}
\usepackage{cancel}
\def\comment#1{}
\def\red#1{\textcolor{red}{#1}}
\def\blue#1{\textcolor{blue}{#1}}

\def\mn#1{\marginpar[\tiny{\red{#1}}]{\tiny{\blu{#1}}}}

\begin{document}

\title{Electromagnetic and gravitational radiation from the coherent oscillation of electron-positron pairs and fields}

\author{Wen-Biao Han$^{a}$ and She-Sheng Xue$^{b}$}
\affiliation{
$^{a}$ Shanghai Astronomical Observatory, 80 Nandan Road, Shanghai, 200030, China.\\
$^{b}$
ICRANet, Piazza della Repubblica 10, I-65122 Pescara, Italy\\
Physics Department and 
ICRA, University of Rome {\it La Sapienza}, Piazzale Aldo Moro 5, I--00185 Rome, Italy.}

\date{Received version \today}

\begin{abstract}
Integrating equations of particle-number and energy-momentum 
conservation and Maxwell field equations, we study the oscillation and drift of electron and positron pairs coherently with fields after these pairs are produced in external electromagnetic fields. From the electric current of oscillating pairs, we obtain the energy spectrum of electromagnetic dipole radiation. This narrow spectrum is so peculiar that the detection of such radiation can identify pair production and oscillation in strong laser fields. We also obtain the energy spectrum of gravitational quadrapole radiation from the energy-momentum tensor of oscillating pairs and fields. Thus, we discuss the
generation of gravitational waves on the basis of rapid development of
strong laser fields.

\end{abstract}
\maketitle

\noindent
{\bf Introduction.}
\hskip0.1cm
Positron and electron pairs are produced from the vacuum in a constant electromagnetic field and the production rate is sizable when the field strength reaches the critical value ($E_c=1.3\times10^{16}\text{V/cm}$); see Refs.~\cite{dunne2006,report}.
To reach this critical value in laboratory, based on recent advanced laser technologies, there are many ongoing  experiments:
x-ray free-electron laser (XFEL) facilities \cite{XFEL}, optical high-intensity laser facilities such as Vulcan or ELI \cite{ELI}, and SLAC E144 using nonlinear Compton scattering \cite{burke1997} for details, see Refs.~\cite{Ringwald,mtb2006,laser_report}. This leads 
to the physics of ultrahigh intensity laser-matter interactions in the critical field \cite{keitel2008}.   

We focus on the backreaction and screening effects of electron and positron
pairs on external electric fields that lead to the phenomenon of plasma
oscillation: electrons and positrons moving back and forth
coherently with alternating electric fields.
In a constant electric field $E_{\rm ext}$, the phenomenon of
plasma oscillations is studied in two frameworks: (1)
the semiclassical QED with a quantized Dirac field and classical
electric field \cite{4,QFT}; and (2) the kinetic description
using the Boltzmann-Vlasov equation (or equations of particle-number and energy-momentum conservation) and the Maxwell equations 
\cite{ion,Matsui1}. 
In Ref.~\cite{4}, two frameworks are discussed. The first framework is semiclassical, where quantized fermion fields $\psi$ satisfy the Dirac equation in an external classical potential $A_\mu$, which satisfies the Maxwell equation coupling to the mean value of charged fermion current. These equations are numerically integrated in (1+1)-dimensional case. The second framework is classical --- the description of particle distribution or density is adopted, 
and the kinetic equation of the Boltzmann-Vlasov for particle density and the Maxwell equation for fields are numerically integrated. The results obtained in two frameworks are in good quantitative agreement \cite{4}, for details, see Refs.~\cite{report}.
In this paper, we adopt the second framework to investigate the plasma oscillations of electron and positron pairs in the (2+1) space-time with the presence of both electric and magnetic fields. We obtain not only the frequencies of plasma oscillations, but also the oscillating pattern of electron and positron pairs in the (2+1) space-time. In addition, we obtain the energy spectra of electromagnetic and gravitational radiation from plasma oscillations.

\section{\bf Plasma oscillation.}

In 1931 Sauter \cite{sauter1931} and four years later Heisenberg and
Euler \cite{he1935} provided a first description of the vacuum properties in
{\it constant} electromagnetic fields. They identified a characteristic scale of strong
field $E_c=
m^2_ec^3/e\hbar$, at which the field energy is sufficient
to create electron positron pairs from the vacuum. In 1951, Schwinger
\cite{schw1951} gave an elegant quantum-field theoretic reformulation
of their result in the spinor and scalar QED framework (see
also \cite{bf1970}). The special attention was given for the presence of magnetic fields \cite{Nikishov1969}. In the configuration of {\it constant} electromagnetic fields, 
the pair-production rate per unit volume is given by
\begin{equation}
\frac{ \Gamma }{V}=\frac{  \alpha   \varepsilon^2}{ \pi^2 }
\sum_{n=1}  \frac{1}{n^2}
\frac{ n\pi\beta / \varepsilon }
{\tanh {n\pi \beta/ \varepsilon}}\exp\left(-\frac{n\pi E_c}{ \varepsilon}\right),
\label{probabilityeh}
\end{equation}
where the two Lorentz invariants
$ \varepsilon $ and $ \beta $ are \begin{eqnarray}\label{fieldinvariant}
\varepsilon \equiv \sqrt{({\mathcal S}^2+{\mathcal P}^2)^{1/2}+ {\mathcal S}}, \quad
\beta  \equiv 
\sqrt{({\mathcal S}^2+{\mathcal P}^2)^{1/2}- {\mathcal S}}.
\end{eqnarray}
In terms of the two
Lorentz invariants, the scalar ${\mathcal S}\equiv ({\bf E}^2-{\bf B}^2)/2=(\varepsilon^2-\beta^2)/2$, and the pseudoscalar ${\mathcal P}={\bf E}\,\cdot {\bf B}=\varepsilon\beta$. In order to focus on studying the phenomenon of plasma oscillations, as a model for quantitative calculations, we postulate an initial configuration of {\it constant} electromagnetic fields: (i) The electric and magnetic fields are perpendicular to each other (${\bf E} \perp {\bf B}$). (ii) Their amplitudes are different ($|{\bf E}|>|{\bf B}|\not=0$) in the laboratory frame, i.e., the rest frame of electron-positron pair production.  For this electromagnetic configuration ${\mathcal P}=0$ and leading term ($n=1$), Eq.~(\ref{probabilityeh}) yields 
\begin{equation}
S=\frac{m_e^4}{4\pi^3}\left(\frac{2{\mathcal S}}{E^2_c}\right)\exp\left[-\frac{\pi E_c}{(2{\mathcal S})^{1/2}}\right], 
\label{srate}
\end{equation}
where the critical field $E_c\equiv m_e^2/e$ and $m_e$ ($-e$) is the electron mass (charge). Note that Eq.~(\ref{srate}) is valid only for ${\mathcal S}> 0$, i.e., $|{\bf E}|>|{\bf B}|$ and ${\bf E} \perp {\bf B}$. In this case $\beta=0$ and $\varepsilon^2=2{\mathcal S}$, Eq.~(\ref{srate}) is equivalent to the case for a purely electric field $E= 2{\mathcal S}$. Equation (\ref{srate}) approaches zero as $|{\bf B}|$ approaches $|{\bf E}|+0^-$. 
As shown below, we have chosen an electric field strength $\bf E$ that is significantly larger than the magnetic one $\bf B$; otherwise, Eq.~(\ref{srate}) would approximately vanish for ${\mathcal S}\approx 0$ and ${\mathcal P}=0$, analogously to the field configuration of a monochromatic laser beam (plane wave ${\mathcal S}={\mathcal P}=0$). We will also discuss the situation in which electromagnetic fields are parallel. It is an important issue for future investigations how initial configurations are dynamically generated from the outset. We adopt $\hbar=c=1$ and Compton units of length $\lambda_C=\hbar/m_ec$, time $\tau_C=\hbar/m_ec^2$, energy scale $m_ec^2$ and critical field strength $E_c$.

In the kinetic description for plasma fluids of positrons ($+$) and electrons ($-$), with single-particle spectra $p_{\pm}^0=({\bf p}_\pm^2+m_e^2)^{1/2}$,
we define the number densities $n_\pm (t,{\bf x})$
and ``averaged''
velocities  ${\bf v}_\pm (t,{\bf x})$ of the fluids:
\begin{align}
n_\pm \equiv  \int  \frac{d^3{\bf p}_\pm}{(2\pi)^3}f_\pm,\quad
{\bf v}_\pm  \equiv  \frac{1}{n_\pm}\int \frac{d^3{\bf p}_\pm}{(2\pi)^3} \left(\frac{{\bf p}_\pm} {p^0_\pm}\right) f_\pm,
\label{meanv}
\end{align}
where $f_\pm=f_\pm(t,{\bf p}_\pm,{\bf x})$ is the distribution function in phase space. The four-velocities of the electron and positron fluids are $U_\pm^{\mu}=\gamma_\pm (1,{\bf v}_\pm)$, the Lorentz factor $\gamma_{\pm}=( 1-|{\bf v}_{\pm}|^{2}) ^{-1/2}$, and the comoving number densities $\bar n_\pm=n_\pm\gamma_\pm^{-1}$.
The collisionless plasma fluid of electrons and positrons coupling to electromagnetic fields is governed
by the equations of particle-number and energy-momentum conservation and the Maxwell equations:
\setlength\arraycolsep{0.5pt}
\begin{align}
&\frac{\partial\left(  \bar{n}_\pm U_\pm^{\mu}\right)  }{\partial
x^{\mu}}
=S;
\quad \frac{\partial T_\pm^{\mu\nu}}{\partial x^{\nu}}
=-F^{\mu}_{~~\sigma}(J_\pm^{\sigma }+J_{\pm\rm pola}^{\sigma
}),\label{contp}\\
&\frac{\partial F^{\mu\nu}}{\partial x^{\nu}}   = -4\pi (J^{\mu}_{\rm cond}+J^{\mu}_{\rm pola}+J^{\mu}_{\rm ext}), \label{me}%
\end{align}
where we have an external electric current $J_{\rm ext}^{\mu} = (\rho_{\rm ext},{\bf J}_{\rm ext})$, electron and positron fluid currents $J_\pm^\mu =\pm e\bar
n_\pm U^\mu_\pm$, and energy-momentum tensors
\begin{align}
T^{\mu\nu}_\pm &= \bar p_\pm g^{\mu\nu}+(\bar p_\pm +\bar \epsilon_\pm)U^\mu_\pm U^\nu_\pm,\quad T^{\mu\nu}_{\rm m}=\sum_\pm T^{\mu\nu}_\pm.
\label{eptensor}
\end{align}
Here the pressure $\bar p_\pm$ and energy density $\bar \epsilon_\pm$ are related by the equation of state $\bar p_\pm=\bar p_\pm(\bar \epsilon_\pm)$ in the fluid comoving frame.
In the laboratory frame, the electron and positron energy density
$p^0_\pm \equiv T^{00}_\pm$ and momentum density $p^i_{\pm} \equiv  T^{i0}_\pm$ are given by $
p^0_\pm =(\bar\epsilon_{\pm}+\bar p_\pm {\bf v}^2_\pm)\gamma^2_{\pm}\, {\rm and}\,\,
{\bf p}_{\pm} =(\bar\epsilon_{\pm}+\bar p_\pm )\gamma^2_{\pm}{\bf v}_\pm .
$
\comment{useful equations: $p^\mu=(p^0_\pm,{\bf p}_{\pm})$ and $T^{ij}_\pm=\bar p_\pm\delta_{ij} +(\bar \epsilon_\pm+\bar p_\pm)\gamma_\pm^2v_{i\pm} v_{j\pm}$.}
The conducting four-current density is
\begin{align}
J_{\rm cond}^{\mu}  &  \equiv e(\bar{n}_+U_+^{\mu} - \bar{n}_- U_-^{\mu}),\quad
\partial_\mu J_{\rm cond}^{\mu} =0,
\label{current}
\end{align}
and the polarized four-current density $J_{\rm pola}^{\mu} = \sum_\pm
J_{\pm\rm pola}^{\mu}$ with $J_{\pm\rm pola}^{\mu}   =
\left(\rho^\pm_{\rm pola}, {\bf J}^\pm_{\rm pola} \right)$
defined by \cite{Matsui1}
\begin{align}
F^\nu_{\,\,\,\,\mu} J_{\pm\rm pola}^{\mu}=\Sigma^\nu_\pm, \quad \Sigma^\nu_\pm =
\int\frac{d^3{\bf p}_\pm}{(2\pi)^3p_\pm^0} p_\pm^\nu {\mathcal A},
\label{pcurrentd}
\end{align}
where ${\mathcal A}$ is related to Eq.~(\ref{srate}) by $S=\int d^3{\bf p}_\pm/[(2\pi)^3p_\pm^0]{\mathcal A}$. $F^{\mu\nu}$ and $T^{\mu\nu}_{\rm em}$ are the field strength and the energy-momentum tensor of electromagnetic fields. 
\comment{
useful equations: 
\begin{align}
T^{\mu\nu}_{\rm em} &= F^\mu_\rho F^{\nu\rho}-\frac{1}{4}\eta^{\mu\nu}F_{\rho\sigma}F^{\rho\sigma}.
\label{ebtensor}
\end{align}
where $\eta^{\mu\nu}=(-1,1,1,1)$, $T^{00}_{\rm em}=({\bf E}^2+{\bf B}^2)/(8\pi)$, $T^{i0}_{\rm em}=({\bf E}\times{\bf B})/(4\pi)$ and $T^{ij}_{\rm em}=-[E_iE_j+B_iB_j-\delta_{ij}({\bf E}^2+{\bf B}^2)/2)]/(4\pi)$.
}

We now assume external electromagnetic fields ${\bf E}_{\rm ext}=E_{\rm ext}\hat {\bf z}$ and ${\bf B}_{\rm ext}=B_{\rm ext}\hat {\bf x}$, where $E_{\rm ext}$ and $B_{\rm ext}$ are constant fields in space and time. 
As will be shown below, in this system, the electron-positron fluid velocities [Eq.~(\ref{meanv})] have $\hat {\bf z}$ and $\hat {\bf y}$ components ${\bf v}_\pm =(v^y_\pm \hat {\bf y} + v^z_\pm \hat {\bf z})$ in the $y-z$ plane, and the total electromagnetic fields are ${\bf E} =E_y \hat {\bf y}+E_z \hat {\bf z}$ and ${\bf B}=B_x\hat {\bf x}$,
which are the superposition of two contributions:
\begin{align}
E_z&=E_{\rm ext}+ \tilde E_z(t,y,z),\quad E_y =\tilde E_y(t,y,z);\nonumber\\
B_x&=B_{\rm ext} + \tilde B_x(t,y,z),\quad B_y =0, \label{tote}
\end{align}
where the space- and time-dependent $\tilde E_{z,y}(t,y,z)$ and $\tilde B_z(t,y,z)$ are the electromagnetic fields created by the motion of electron and positron pairs.

We adopt the approximations $\bar p_\pm\approx 0$, $\bar\epsilon_\pm\approx m_e \bar n_\pm$, and $ \epsilon_\pm=\bar \epsilon_\pm\gamma_\pm^2$ when the pair number density is not very large for $E\simeq E_c$. Using Eqs.~(\ref{meanv}) and (\ref{pcurrentd}), we obtain
\comment{these equations are useful, we drop them for limiting pages
\begin{align}
E_yJ^0_{\pm\rm pola} + B_xJ^z_{\pm\rm pola} &\approx  m_e\gamma_\pm v^y_\pm S,\nonumber\\
E_zJ^0_{\pm\rm pola} - B_xJ^y_{\pm\rm pola} &\approx  m_e\gamma_\pm v^z_\pm S,\nonumber\\
E_yJ^y_{\pm\rm pola} + E_zJ^z_{\pm\rm pola} &\approx  m_e\gamma_\pm  S;
\label{pcurrent0}
\end{align}
and
}
\begin{align}
J^{z,y}_{\rm pola}  &\approx \frac{E_{z,y}}{E^2}\left(m_e\gamma_\pm S\right),\,\,
J^0_{\pm\rm pola} \approx \frac{v^z_\pm E_z+v^y_\pm E_y}{E^2}m_e\gamma_\pm S,
\nonumber
\end{align}
where $E^2=E_z^2+E_y^2$. 
The total electric current and charge densities of the  electron-positron fluid are composed by Eqs.~(\ref{current}) and (\ref{pcurrentd}) as
\begin{align}
J_z &= e_+ n_{+}v^z_{+} + e_-n_{-}v^z_{-}
+J^z_{+\rm pola}+J^z_{-\rm pola},\label{tcur}
\end{align}
$J_y=J_z(z\rightarrow y)$ and $\rho = \sum_\pm (e_\pm n_{\pm}+J^0_{\pm\rm pola})$,
where the positron and electron charge $e_\pm \equiv \pm e$. 

It turns out to be a $(1+2)$-dimensional problem in space-time coordinates $(t,y,z)$. Equations (\ref{contp}) and (\ref{me}) are reduced to
(i) the particle-number and energy conservation,
\begin{align}
\frac{\partial n_\pm}{\partial t}&\!+\! \frac{\partial n_\pm
v^z_\pm}{\partial z}\!+\!\frac{\partial n_\pm
v^y_\pm}{\partial y}=S,\label{nudot}\\
\frac{\partial\epsilon_\pm}{\partial t}&\!+\!
\frac{\partial p^z_\pm}{\partial z}\!+\!\frac{\partial p^y_\pm}{\partial y}= e_\pm n_\pm v^z_\pm E_z \!+\! e_\pm n_\pm v^y_\pm E_y\!+\!m_e\gamma_\pm S;
\nonumber
\end{align}
(ii) the momentum conservation,
\begin{align}
\!\!\frac{\partial p^z_\pm}{\partial t}&\!+\!\frac{\partial p^z_\pm
v^z_\pm}{\partial z} \!+\!\frac{\partial p^z_\pm
v^z_\pm}{\partial y}=e_\pm n_\pm E_z \!-\! e_\pm n_\pm v^y_\pm B_x\!+\! E_zJ^0_{\pm\rm pola}
\nonumber
\end{align}
with $(z\leftrightarrow y, B_x\rightarrow -B_x)$;
(iii) Maxwell equations $\nabla\cdot {\bf E}= 4\pi \rho$,  $\nabla\cdot {\bf B}=0$,
\begin{align}
\frac{\partial \tilde E_z}{\partial t}+\frac{\partial \tilde B_x}{\partial y} =-4\pi J_z,
\quad
\frac{\partial \tilde E_z}{\partial y}-\frac{\partial \tilde E_y}{\partial z}=-\frac{\partial \tilde B_x}{\partial t},\label{Ediv-d}
\end{align}
with $(z\leftrightarrow y, B_x\rightarrow -B_x)$.
The pair-production rate [Eq.~(\ref{srate})] can be approximately
used for varying electromagnetic fields [Eq.~(\ref{tote})],
provided $\tilde E(t,y,z)$ and $\tilde B(t,y,z)$ created by
electron-positron pair oscillations vary very slowly compared with the rate of electron-positron pair productions ${\mathcal O}(m_ec^2/\hbar)$.
This is justified if
the inverse adiabaticity
parameter \cite{Brezin}
$
\eta=\frac{m_e}{\omega_p}\frac{E}{E_{c}}\gg 1,
$
where $\omega_p$ is the frequency of plasma oscillations.

We are in the position of numerically integrating the basic equations (\ref{nudot}) and (\ref{Ediv-d}). The initial conditions
$(t=0)$ are given by the constant electromagnetic fields $E_z=E_{\rm ext}$ and $B_x=B_{\rm ext}$. To simplify numerical integrations, we assume the $(z\!-\!y)$ homogeneity that the electron-positron fluid quantities and electromagnetic fields are independent of $y$ and $z$. As a result,
Eqs.~(\ref{nudot}) and (\ref{Ediv-d}) are reduced to ordinary differential equations, and  Eq.~(\ref{Ediv-d}) leads to $\tilde B_x=0$, i.e., the magnetic field $B_x$ of Eq.~(\ref{tote}) is a constant in space and time.
\comment{
In this system, except electron and positron velocities are in opposite directions, their quantities, including electric currents, are the same. Therefore, we only need to solve the equations for positrons and  $(E,B)$-fields. The positron current represents the electric current
and we hereafter omit its subscript $``+''$. In the following, we present our numerical studies of the plasma oscillations.
}
The initial condition $E_y=0$ leads to the solution $\tilde E_y=0$ and $J_y=0$ for $t\not=0$, because $v^z_-=-v^z_+$ and $v^y_-=v^y_+>0$. This is verified in the following numerical calculations.

\begin{figure}
\begin{center}
\includegraphics[height=1.25in]{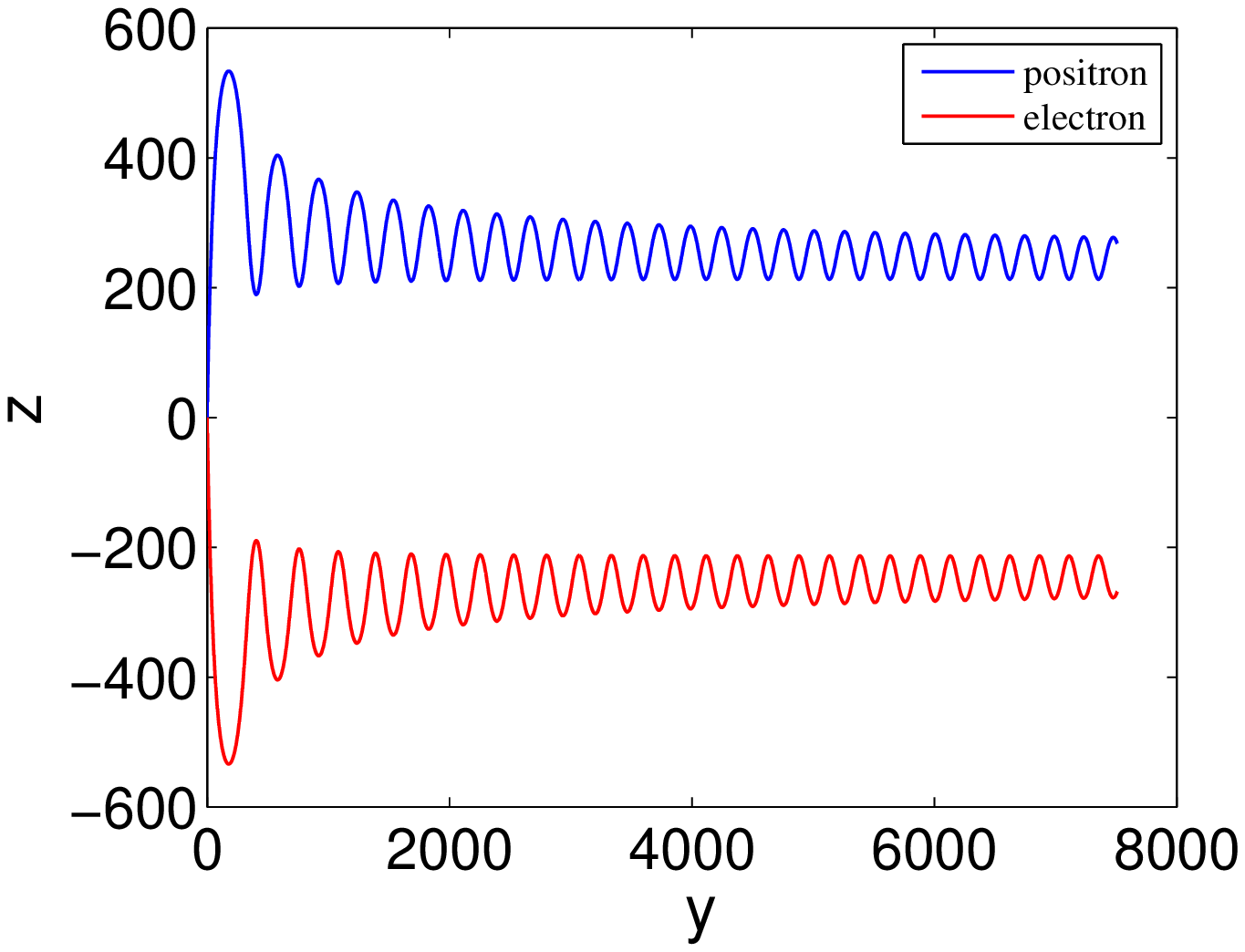}
\includegraphics[height=1.25in]{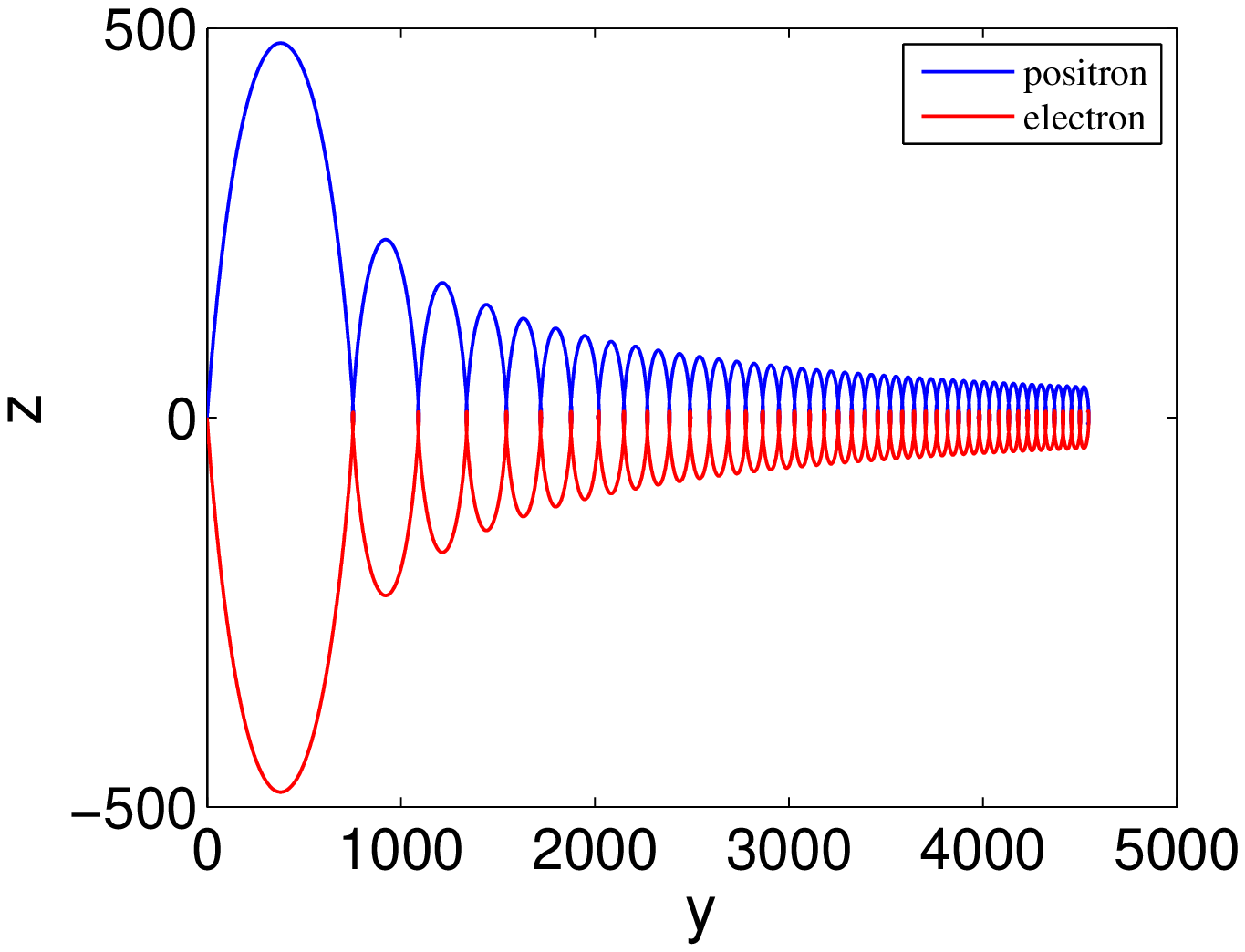}
\includegraphics[height=1.25in]{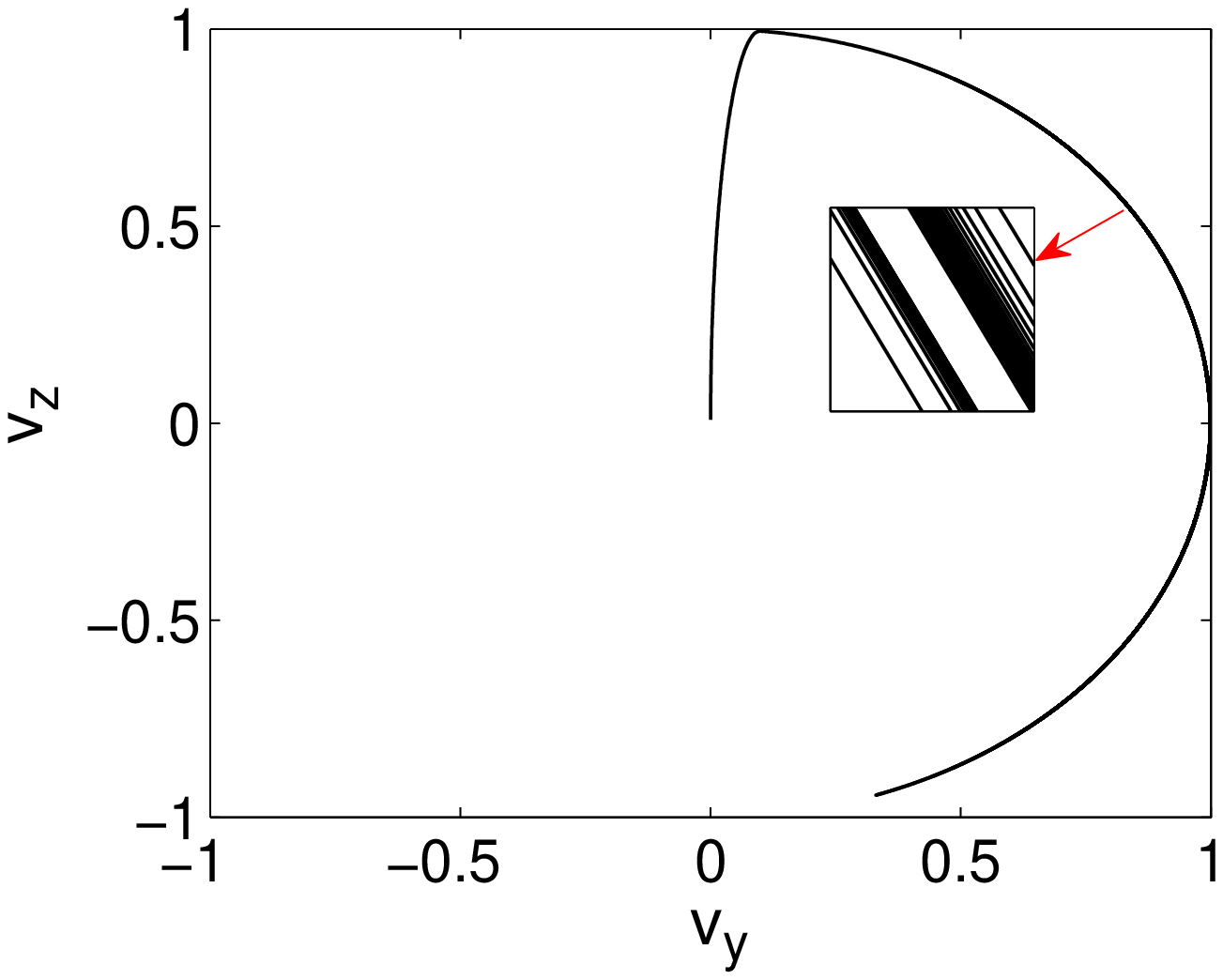}
\includegraphics[height=1.25in]{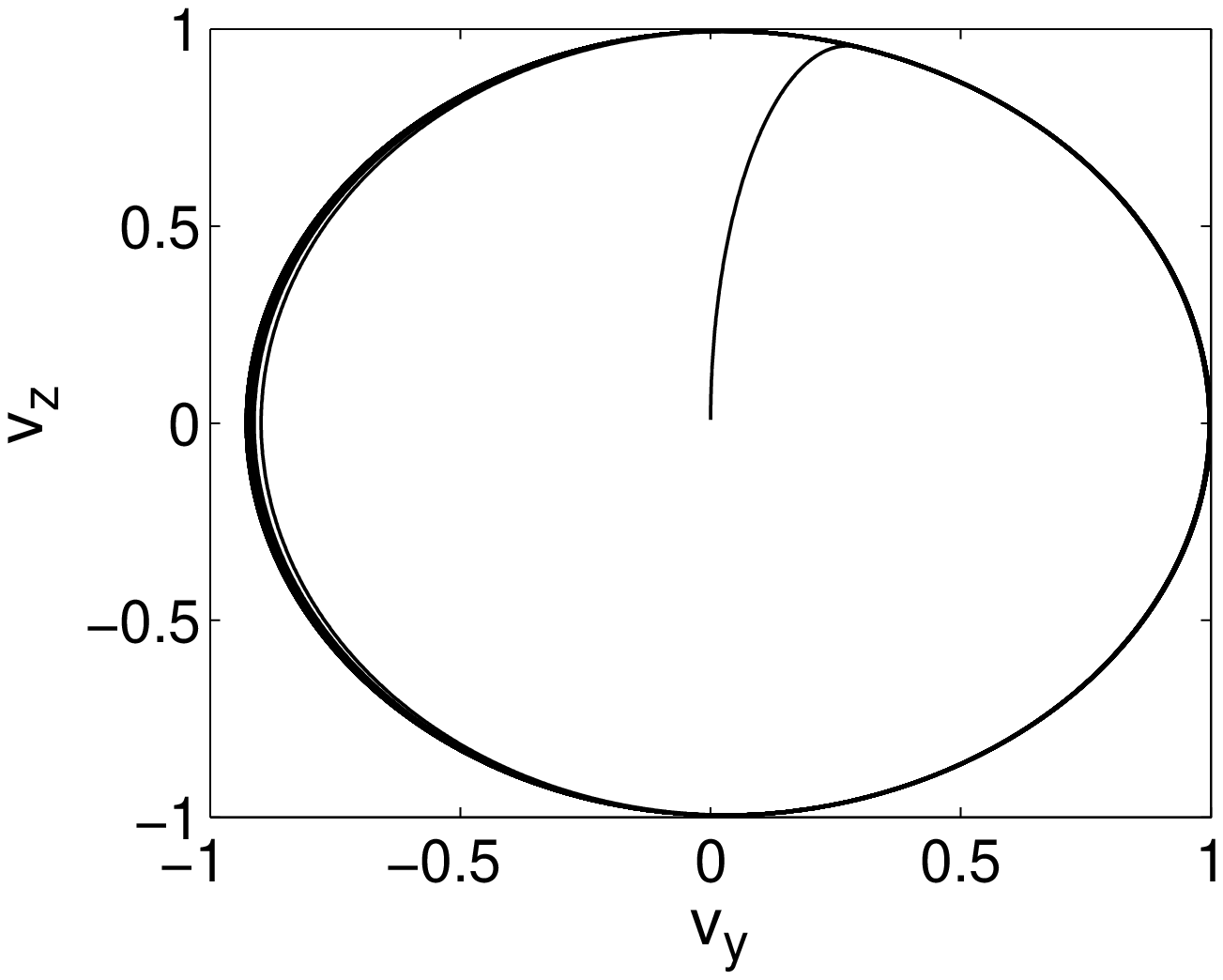}
\includegraphics[height=1.25in]{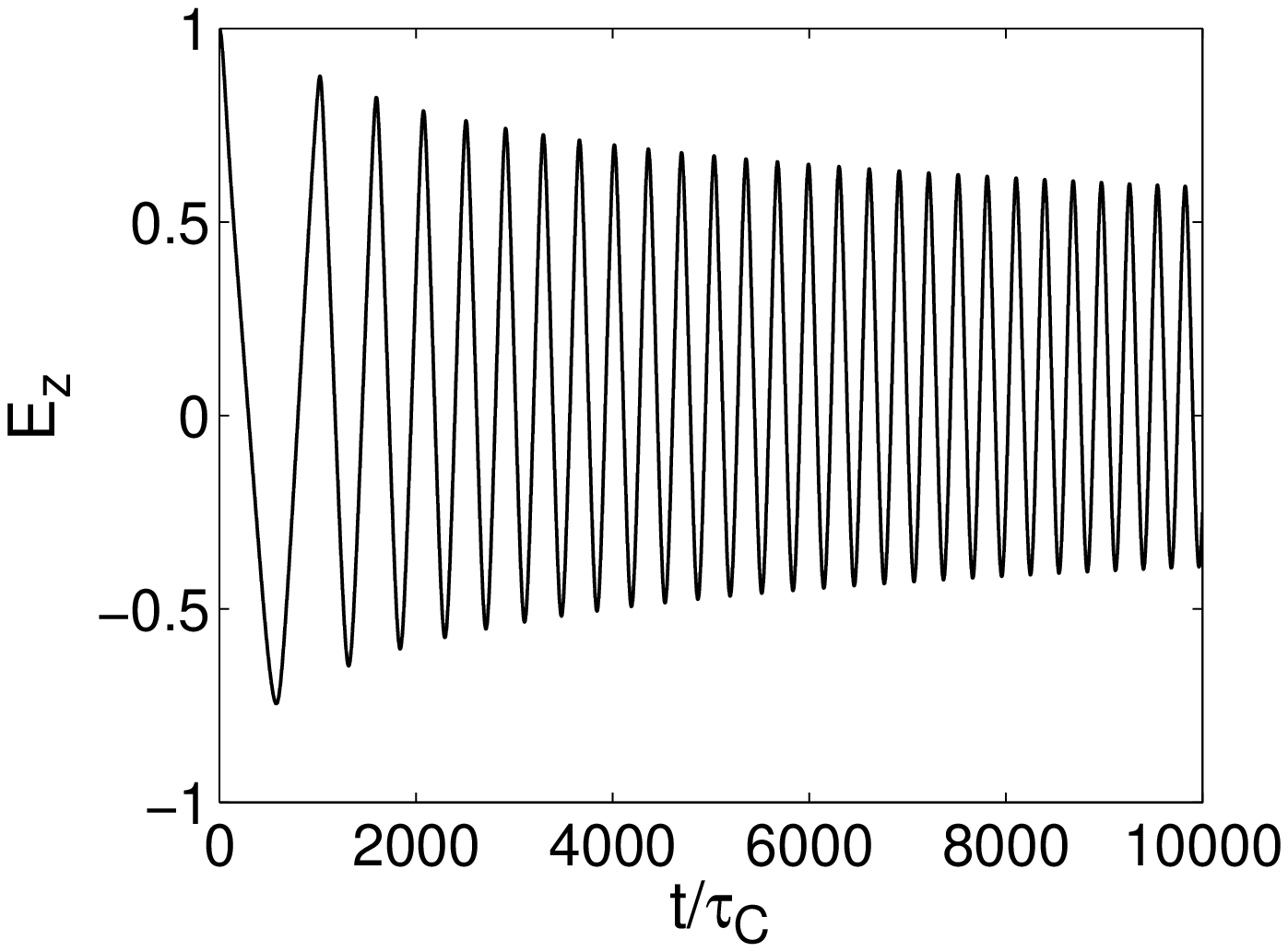}
\includegraphics[height=1.25in]{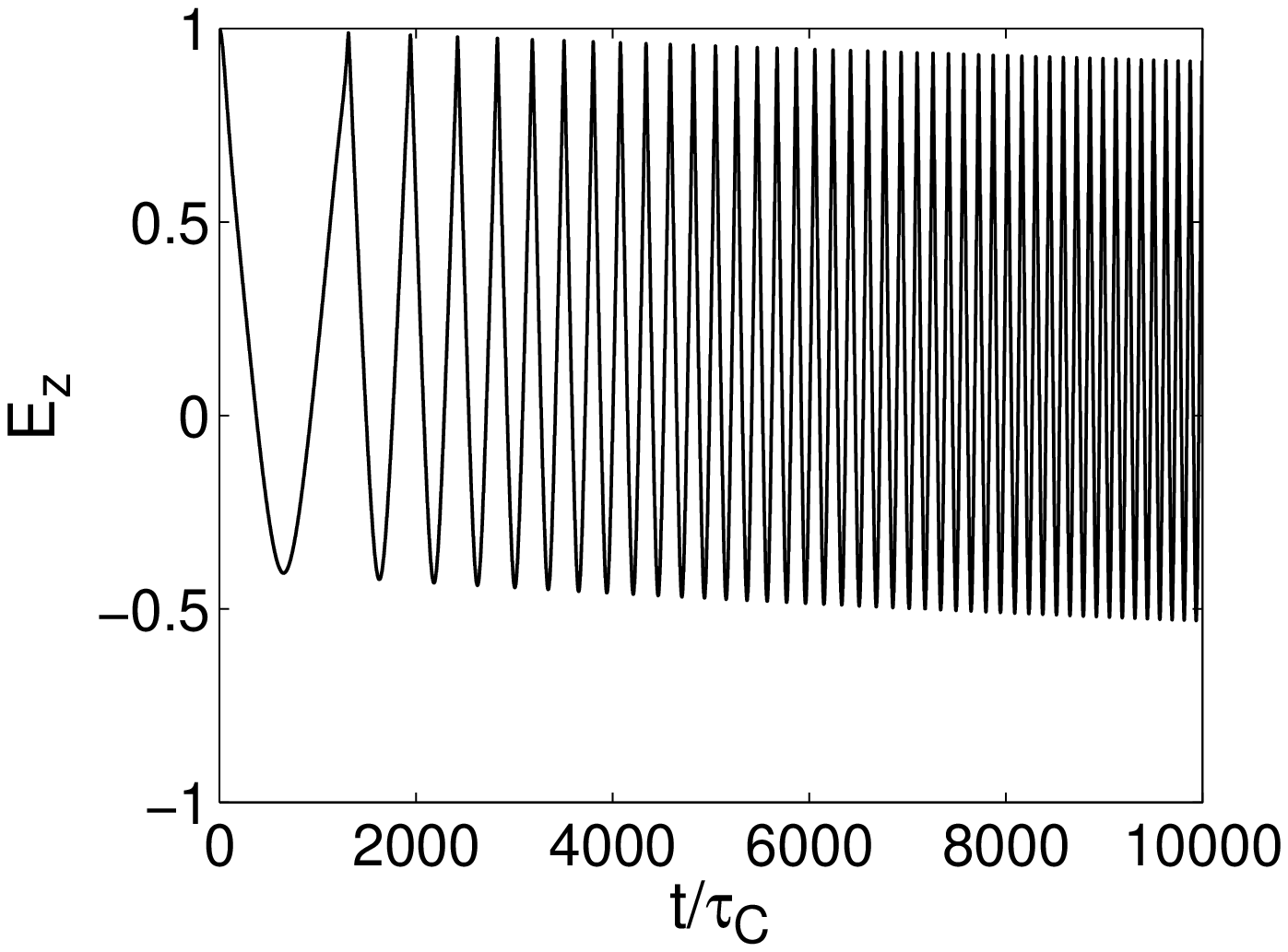}
\caption{With initial conditions $E_z=E_c$, $B_x=0.1E_c$ (left) and $B_x=0.3E_c$ (right), we plot the electron and positron trajectories and velocities $v_y$ vs $v_z$ and the electric field $E_z$ vs the time $t$ from $t=0$ to $t= 10^4$.} \label{figm}
\end{center}
\end{figure}

To illustrate the plasma oscillations of pairs and fields, we consider two cases: (i) $E_{\rm ext}=E_c$ and $B_{\rm ext}=0.1\,E_c$; (ii) $E_{\rm ext}=E_c$ and $B_{\rm ext}=0.3\,E_c$. Due to the presence of the magnetic field $B_x$, pairs are not only oscillating up and down in the $\hat {\bf z}$ direction, as first shown in Ref.~\cite{4}, but they also move in the $\hat {\bf y}$ direction.  In Fig.~\ref{figm}, we show the trajectory and velocity of pairs produced at $z=y=0$ and $t=0$. When $B_x\not=0$ and $dv_y \sim ev_zB_x dt$, $v_y$ increases for $v_z >0$ and decreases for $v_z <0$.
In the case of $B_x$ being small enough compared with $E_z$, $v_y$ does not change its sign (see Fig.~\ref{figm}, $B_x=0.1E_c$) in the period of one circle oscillation in the ${\bf \hat z}$ direction; therefore pairs move forward in the ${\bf \hat y}$ direction. When $B_x=0.3\, E_c$, $v_y$ changes its sign (see Fig.~\ref{figm}, $B_x=0.3E_c$); therefore pairs also oscillate back and forth, while they are moving in the ${\bf \hat y}$ direction. In contrast to the case $B_x=0$, the negative $E_z$ amplitude is smaller than the positive $E_z$ amplitude (see Fig.~\ref{figm}). The reason is that $v_y$ increases in the phase of positive decreasing $v_z$ when $E_z<0$; i.e., the electric energy goes to the kinetic energy of the motion in the $\hat {\bf y}$ direction.

\begin{figure}
\begin{center}
\includegraphics[height=1.25in]{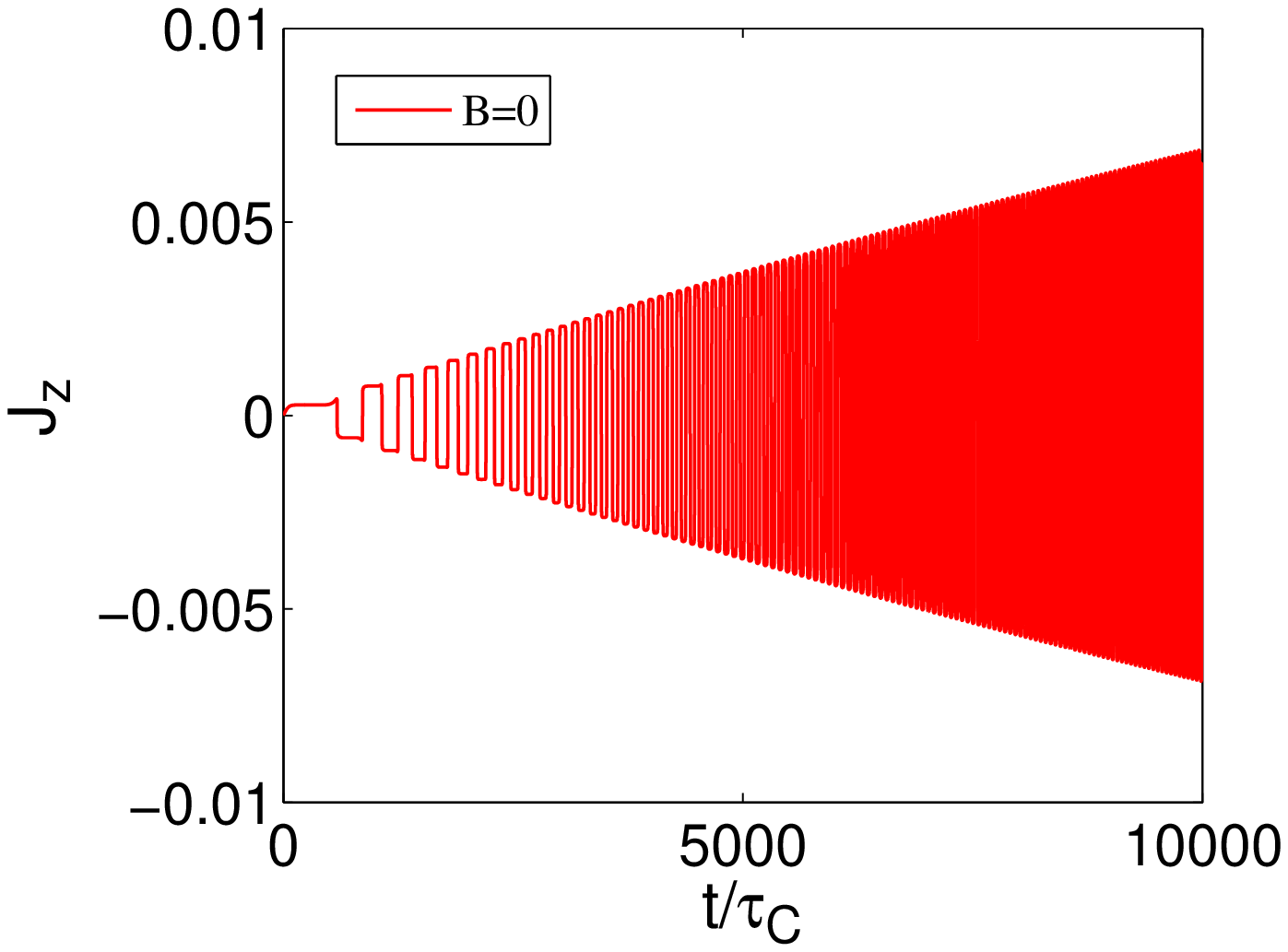}
\includegraphics[height=1.25in]{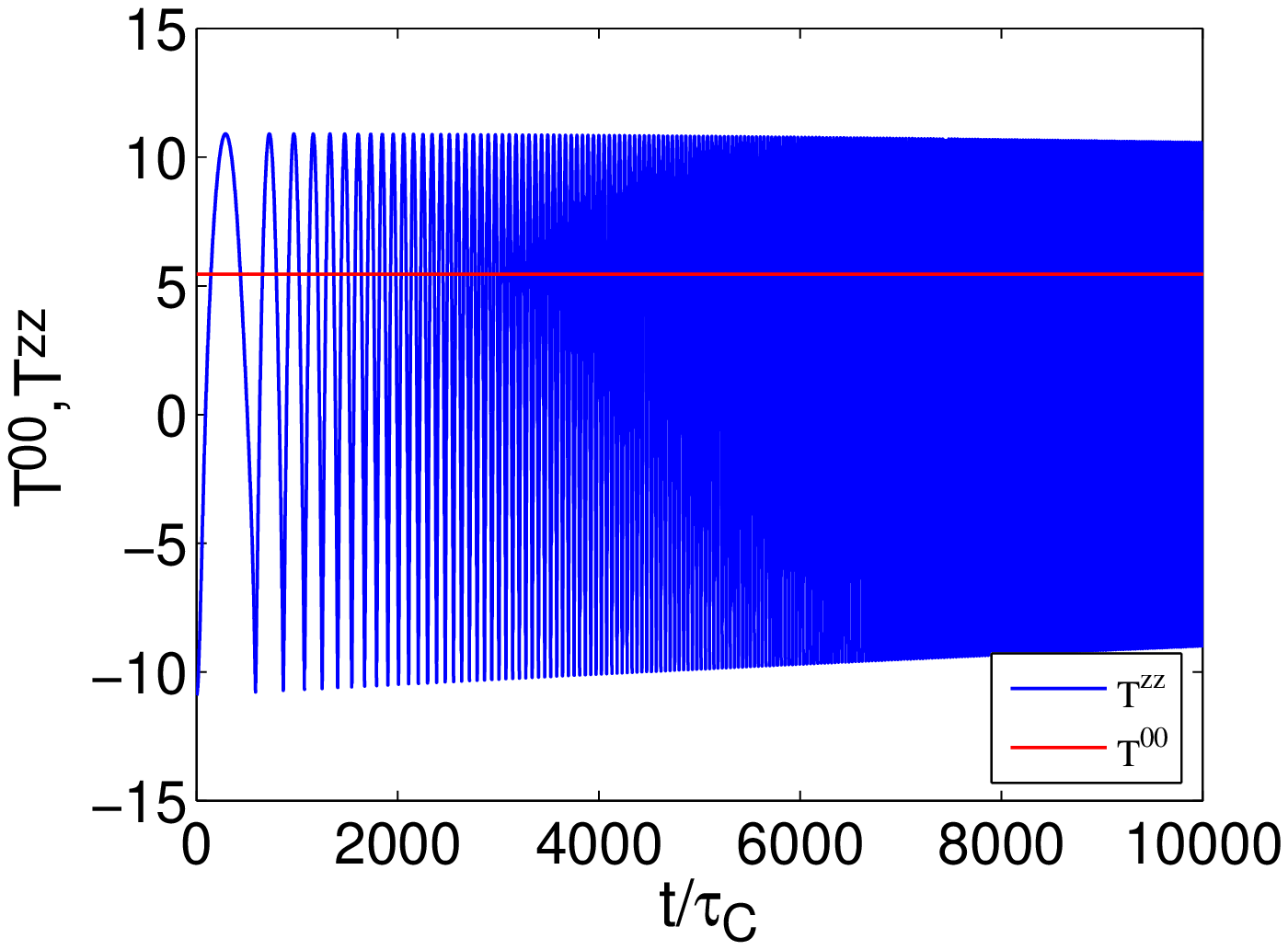}
\caption{$E_z=E_c$ and $B_x=0.0$. The charged current density $J_z$ vs time $t$ (left). The total energy-momentum tensor $T^{00}$ and $T^{zz}$ vs time $t$ (right).} \label{figT}
\end{center}
\end{figure}

In Fig.~\ref{figT}, we plot 
(i) the electric current density of pairs $J_z$ as a function of the time, which is the source of electromagnetic radiation, and
(ii)the total energy-momentum tensor of pairs and fields $T^{\mu\nu}=T^{\mu\nu}_{\rm m}+T^{\mu\nu}_{\rm em}$ as functions of the time, which are the sources of gravitational radiation.

\begin{figure}
\begin{center}
\includegraphics[height=1.25in]{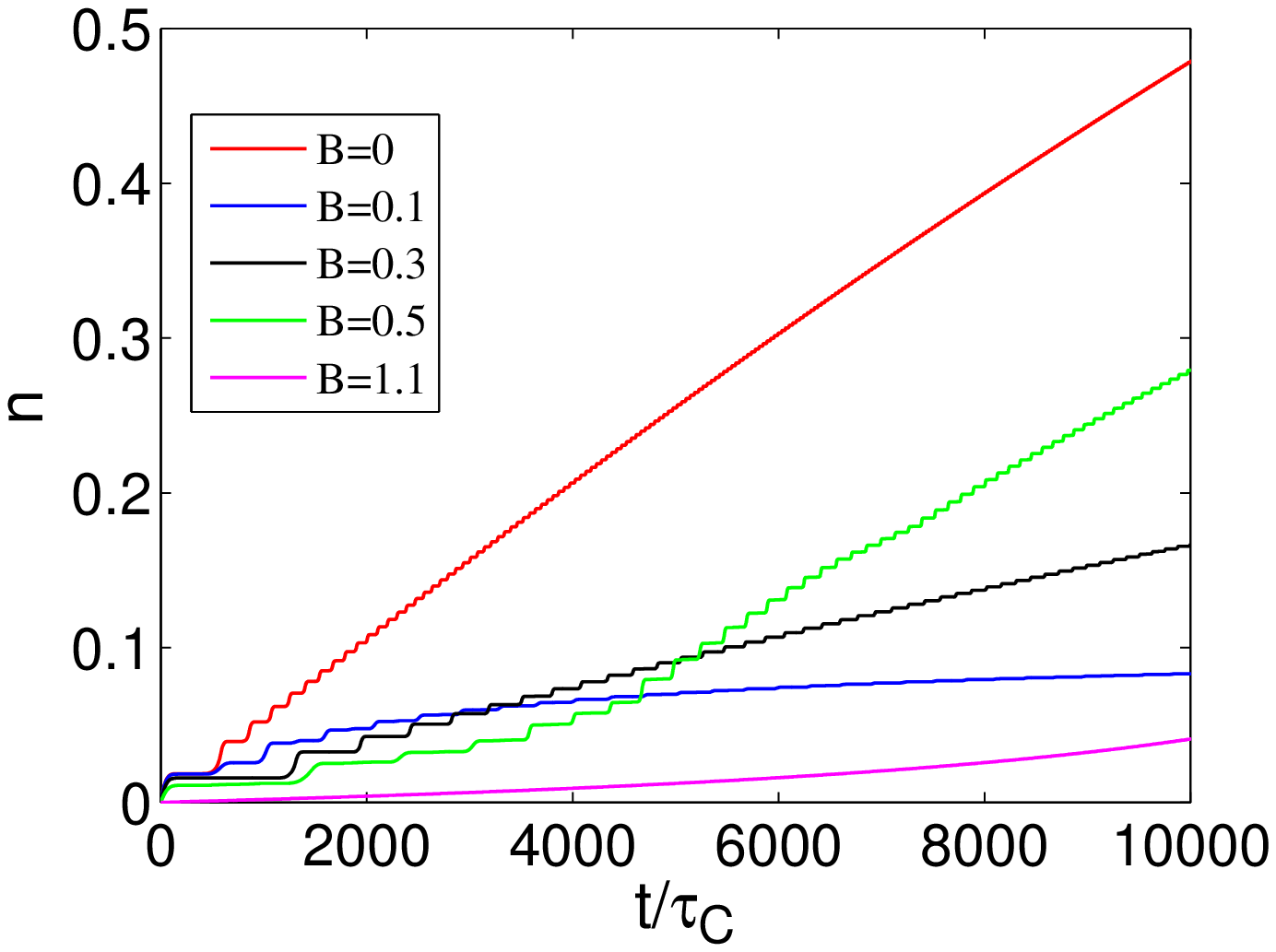}
\includegraphics[height=1.25in]{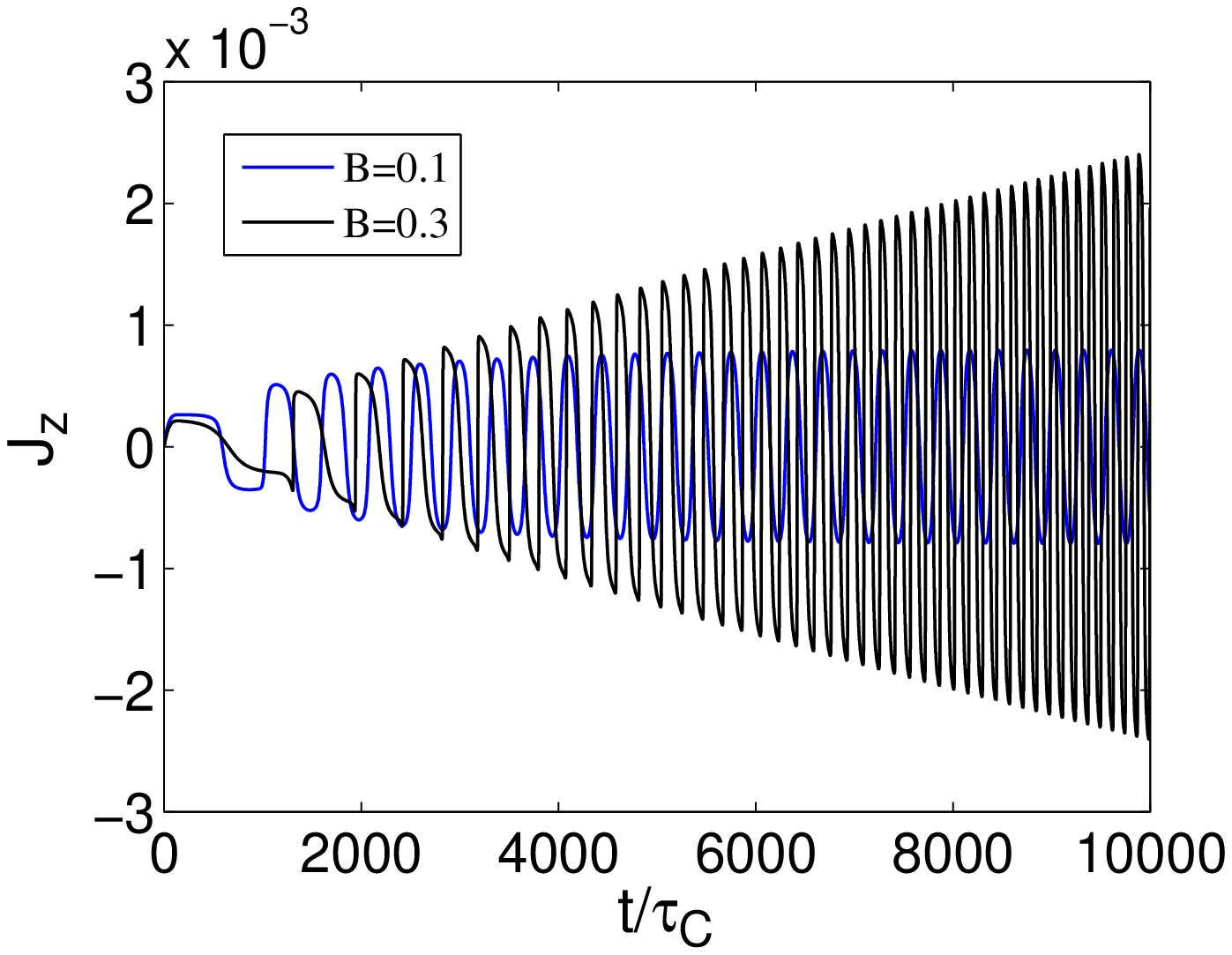}
\caption{The pair number density $n_\pm$ and current density $J_z$ vs the time $t$ for $E_z=E_c$ and different $B_{x}$ field values.}
\label{fign}
\end{center}
\end{figure}

Before ending this section, we would like to present some discussions on the role of magnetic fields. In the particular initial configuration of fields ${\bf E} \perp {\bf B}$ and $|{\bf E}|>|{\bf B}|$
considered in this paper, by integrating Eqs.~(\ref{srate}), (\ref{contp}) and (\ref{me}), we show the oscillating electric field strength (see Fig.~\ref{figm}), and the number and current densities of pairs (see Fig.~\ref{fign}) are suppressed by magnetic fields, compared with their counterparts in the absence of magnetic fields. However, we cannot conclude that such magnetic suppression is generally true. For example, when electromagnetic fields are parallel (${\bf E}\times{\bf B}=0$ and $|{\bf E}|>|{\bf 
B}|$), Eq.~(\ref{probabilityeh}) yields (see Ref.~\cite{Nikishov1969})
\begin{equation}
\frac{\Gamma}{V}\simeq
\frac{\alpha |{\bf B}||{\bf E}|}{\pi}\coth\left(\frac{\pi |{\bf B}|}{ |{\bf E}|}\right)
\exp\left(-\frac{\pi E_c}{|{\bf E}|}\right),
\label{wkbehfermion1}
\end{equation}
indicating that the pair-production rate receives an enhancement $(\pi |{\bf B}|/|{\bf E}|) \coth (\pi |{\bf B}|/|{\bf E}|)$ to the prefactor, compared with the rate in the absence of magnetic fields \cite{Nikishov1969,supp_mag} (see also \cite{dunne2006,report}). 
It is worthwhile to study the phenomenon of plasma oscillations by numerically integrating Eqs.~(\ref{contp}), (\ref{me}) and (\ref{wkbehfermion1}) consistently with the initial configuration of parallel electromagnetic fields \cite{future}. 

\vskip0.1cm \noindent{\bf Electromagnetic and gravitational radiation.}
We attempt to study electromagnetic and gravitational radiation generated, respectively,  by the electric current and energy-momentum tensor of pairs and fields. Suppose that we observe this radiation in the {\it wave zone}; that is, at distances much larger than the dimension ${\mathcal R}$ of the plasma oscillations, and also much larger than $\omega {\mathcal R}^2$ and $1/\omega$, where $\omega$ is the typical frequency of radiation.

For definiteness we think of the electric current and energy-momentum tensor of the plasma oscillations occurring
in the volume ${\mathcal
V}$ and for a finite interval of time ${\mathcal T}$, so that the total energy radiated is finite. Thus, the electromagnetic energy radiated per unit solid angle per frequency interval is given by \cite{Jackson}
\begin{align}
\!\!\frac{d^2{\mathcal E}^{^{\rm em}}}{d\omega d\Omega}
=2\left|\int_{\mathcal
V} d^3x' \int_{\mathcal T} dt' e^{i\omega t'-i{\bf k}{\bf x}'}\left[\frac{\partial J_z({\bf x}',t')}{\partial t'}\right]\right|^2.
\label{intem}
\end{align}
The gravitational energy radiated per unit solid angle per frequency interval is then given by \cite{Weinberg1972}
\begin{align}
&\frac{d^2{\mathcal E}^{^{\rm grav}}}{d\omega d\Omega} = 2G\omega^2\Big[T^{\mu\nu *}({\bf k}, \omega)T_{\mu\nu }({\bf k},\omega)-\frac{1}{2}|T^\nu_{\,\,\,\nu }({\bf k}, \omega)|^2\Big],\nonumber\\
&T_{\mu\nu }({\bf k}, \omega) =
\int_{\mathcal
V} d^3x' \int_{\mathcal T} dt' \,\, T_{\mu\nu }({\bf x}', t')e^{i\omega t'-i{\bf k}{\bf x}'}\label{intg}
\end{align}
where $|{\bf k}|=\omega$ and $T^{\mu\nu }({\bf x}', t')=T^{\mu\nu }_{\rm m}({\bf x}', t')+T_{\rm em}^{\mu\nu }({\bf x}', t')$. We consider $\omega {\mathcal R}\ll 1$ and $e^{-i{\bf k}{\bf x}'}\approx 1$ for dipole electromagnetic radiation in Eq.~(\ref{intem}), and for quadrapole gravitational radiation in Eq.~(\ref{intg}). In the calculations of Eq.~(\ref{intg}), we set $B_{\rm ext}=0$ and $E_{\rm ext}=E_c$, and then the nonvanishing components are $T^{00}=T^{00}_{\rm m}+T^{00}_{\rm em}$ and $T^{zz}=T^{zz}_{\rm m}+T^{zz}_{\rm em}$.
Using the approximation of spatial homogeneity in Eqs.~(\ref{intem}) and (\ref{intg}), we can factorize out the volume ${\mathcal V}=\int_{\mathcal V} d^3x'$, in which the total energy density $T^{00}=T^{00}_{\rm m}+T^{00}_{\rm em}=E_{\rm ext}^2/(8\pi)$ is conserved (see Fig.~\ref{figT}).

\begin{figure}
\begin{center}
\includegraphics[height=1.25in]{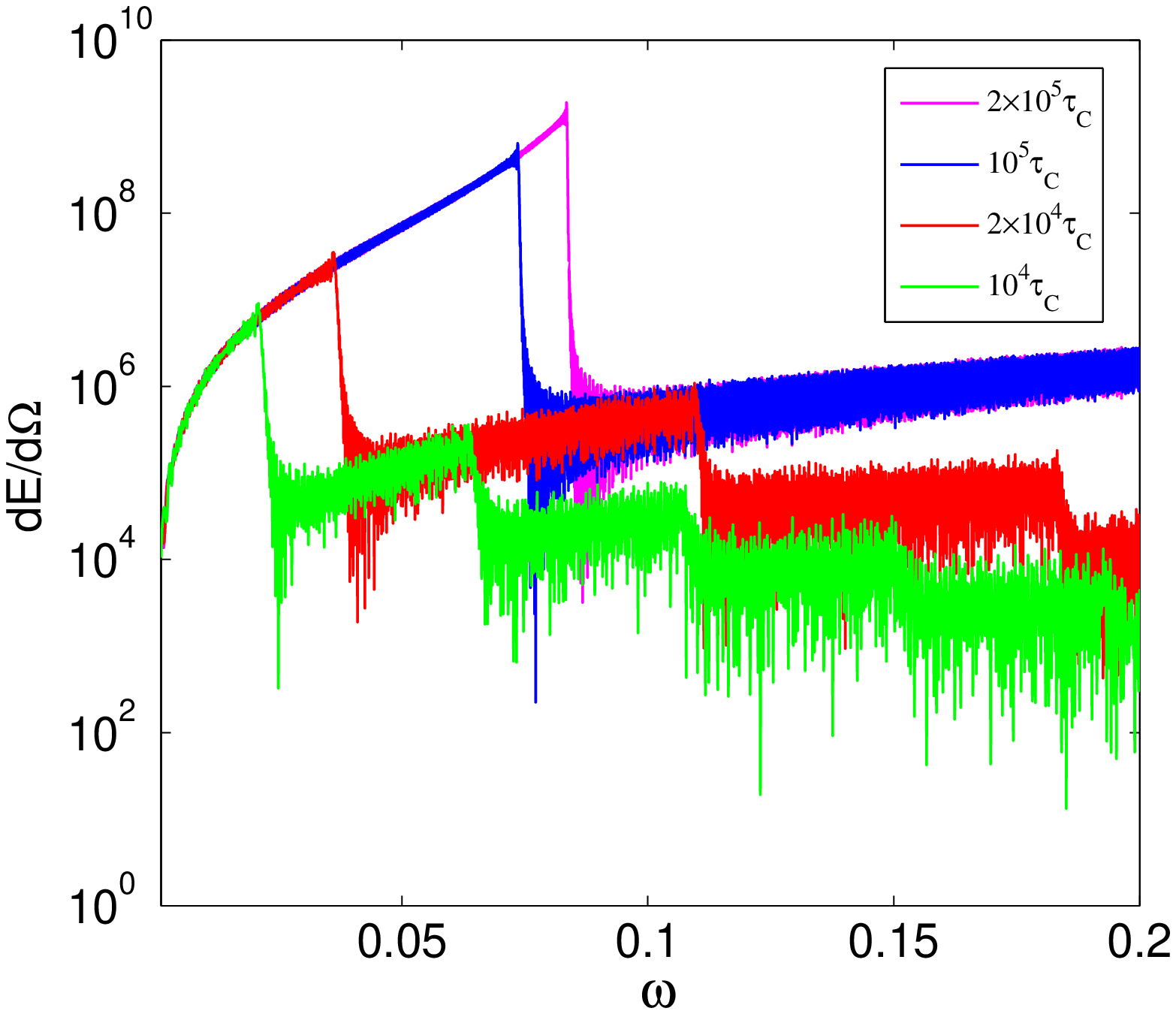}
\includegraphics[height=1.25in]{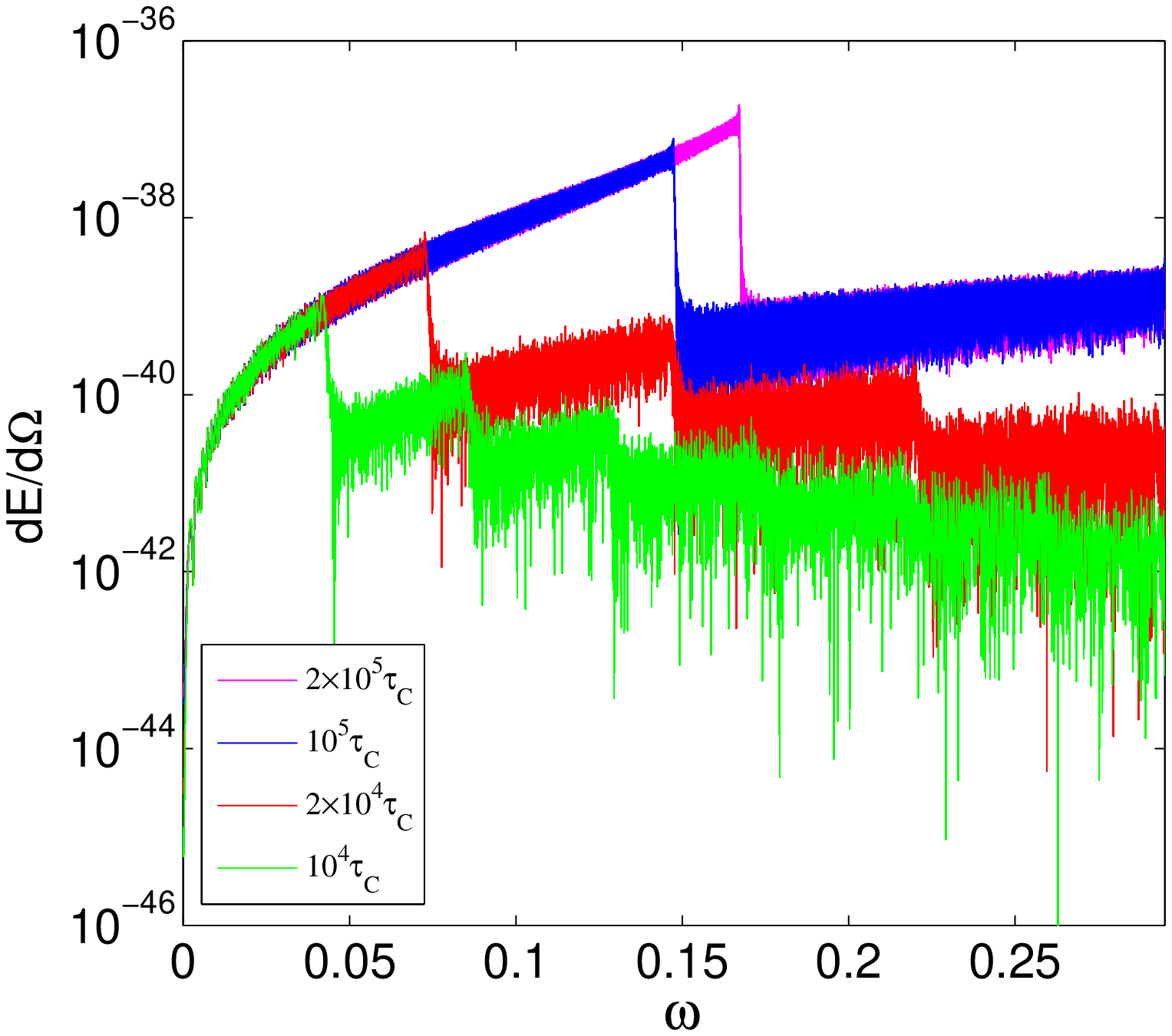}
\caption{$E_z=E_c$ and $B_x=0.0$. By factoring ${\mathcal V}^2$ out, the electromagnetic radiation (left) of Eq.~(\ref{intem}) and the gravitational radiation (right) of Eq.~(\ref{intg}) are plotted as functions of the frequency $\omega$, for different ${\mathcal T}$ values .} \label{figR}
\end{center}
\end{figure}

Let ${\mathcal T}$ and ${\mathcal V}$ also be the time and volume of strong fields $E_{\rm ext}\gtrsim E_c$ created by coherent laser beams. Selecting different ${\mathcal T}$ values, in Fig.~\ref{figR} we plot the electromagnetic and gravitational radiation spectra (\ref{intem}) and (\ref{intg}) with ${\mathcal V}^2$ factored out. These two energy spectra are narrow, and the locations ($\omega_{\rm peak}$) of their peaks are related to the coherent oscillation frequency ($\omega_p$) of pairs and fields, which depend on ${\mathcal T}$ and $E_{\rm ext}$ (see Ref.~\cite{Han1}). The peculiar energy spectrum of electromagnetic radiation is clearly distinguishable from the energy spectra of
the bremsstrahlung radiation, electron-positron annihilation and other possible background events. Therefore, it is sensible and distinctive to detect such peculiar radiative signatures to identify the production and
oscillation of electron-positron pairs in strong laser fields. 
As shown in Fig.~\ref{figR}, gravitational radiation is much smaller than the electromagnetic one for the reason that the gravitational coupling $Gm_e^2=2.5\times 10^{-45}$ is much smaller than the electromagnetic coupling $e^2=1/137$. 
\comment{For example, the pair density $n\sim 10^{31}{\rm cm}^{-3}$ for $E_{\rm ext} \sim E_c$ (see Fig.~\ref{fign}), from our results (see Fig.~\ref{figR}) the intensities of electromagnetic and gravitational radiation are given by
\begin{eqnarray}
I_{\rm grav}&\approx& 1.56\cdot 10^{-29}\,\,{\mathcal V}^2({\rm ergs/sec})=1.56\cdot 10^{-11}({\rm ergs/sec});\nonumber\\
I_{\rm em}&\approx& 1.56\cdot 10^{17}\,\,{\mathcal V}^2({\rm ergs/sec})=1.56\cdot 10^{35}({\rm ergs/sec}).
\label{ints}
\end{eqnarray}
where ${\mathcal V} = (3.86\cdot 10^{-8} {\rm cm}/\lambda_C)^3$.} 
In order to achieve a sizable radiation intensity from the plasma oscillation, the volume ${\mathcal V}$ of oscillating pairs and strong electric fields should be large enough and/or the strength of strong fields should be enhanced ($E_{\rm ext} \gtrsim E_c$) to increase the pair density. 
It is worthwhile to point out that Fig.~\ref{figR} shows the numerical results of Eqs.~(\ref{intem}) and (\ref{intg}) being consistent with the approximate relation between Eqs.~(\ref{intem}) and (\ref{intg}) in the ultrarelativistic limit of charged particles moving in external electromagnetic fields \cite{ritus1968}.
 
Gravitational waves from an inflationary cosmos \cite{cosmos} are in the high frequency band $(10^{8}-10^{11}$Hz).
Gravitational waves originating from some sources in ground laboratories are also in this frequency band, and several proposals have been made to detect high-frequency gravitational waves up to $5$ GHz  \cite{high1}.  Gravitational waves generated from  the high-energy particle beam \cite{ritus1968,hep-grav}
in the ground experiments of the Stanford Linear
Collider and LHC
have much higher frequencies of ${\mathcal O}(10^{23})$ Hz. The frequency of gravitational wave discussed here is ${\mathcal O}(10^{19-20})$ Hz, i.e., ${\mathcal O}(10^{0-1})$KeV,
or sub nanometer ${\mathcal O}(10^{-(9-10)})$ cm. 
It is not clear
whether such gravitational waves could ever be detected, or have observable effects. One would have to build an atom-sized gravitational wave detector to response incoming gravitational wave with such high frequencies (for some more details, see Ref.~\cite{weiton2010}) 
\comment{We try to make some speculations that If would have been realized, possible candidates for such a detector would be highly coherent systems of particles and fields, for instance, a neutral electron-proton plasma of frequency $\omega_{pe}\approx\sqrt{4\pi e^2 n_e/m_e}\sim 10^{19}$Hz when the electron density $n_e\sim 10^{30}/{\rm cm^3}$, and/or a system of electron-positron plasma oscillations discussed in this paper.} 


To end this paper, we remark again that
the intensity of electromagnetic
radiation emitted by the plasma oscillations is tens of orders of magnitude larger than their gravitational radiation; therefore any detectable signal is enormously more likely to result from the electromagnetic interaction. The prospect of detecting gravitational radiation of such ultrahigh frequencies looks dim. Nevertheless, our theoretical investigation of the gravitational radiation from the electron-positron plasma oscillation would be useful for the study of gravitational radiation emitted from particles and antiparticles in the very early Universe.

\comment{
To end this paper, we remark that 
since such gravitational waves are emitted from a highly coherent system of particles and fields created by the advanced laser technique, it might be possible to set up two systems that emit and receive coherent gravitational waves, leading to detectable signals of resonance and/or interference effects. 
}


\noindent{\bf Acknowledgements:} Wen-Biao Han is supported by NSFC Grant No.11273045.


\begin{thebibliography}{99}
\bibitem{future}
We are planning to do this investigation to understand the role of magnetic fields on the pair-production and plasma oscillations.
\end{thebibliography}
\end{document}